\documentclass{optica-article}

\journal{opticajournal} 

\articletype{Research Article}

\usepackage{lineno}

\begin{document}

\title{Impact of interference between two infrared pulses driving high harmonic generation}

\author{Sarang Dev Ganeshamandiram,\authormark{1} Jahanzeb Muhammad,\authormark{1} Marvin Schmoll,\authormark{1} Ronak Shah,\authormark{1} Frank Stienkemeier,\authormark{1} Giuseppe Sansone,\authormark{1} and Lukas Bruder \authormark{1,*}}

\address{\authormark{1}Institute of Physics, University of Freiburg, Hermann-Herder-Str. 3, 79104 Freiburg, Germany}

\email{\authormark{*}lukas.bruder@physik.uni-freiburg.de} 


\begin{abstract*} 
Extreme ultraviolet (XUV) interferometry is technically challenging to implement. 
One approach to generating interference between two XUV pulses relies on driving high-harmonic generation in a gas jet with two collinearly overlapping infrared laser pulses. 
We investigate this scheme through a combined experimental and theoretical study, with particular emphasis on the regime of temporal overlap between the driving pulses. 
A special phase-modulation interferometry technique is implemented to increase the sensitivity for the comprehensive mapping of the strong-field induced high-order nonlinear response. 
We find that the dynamics arising from the interference of the two electric fields can be adequately described by the non-perturbative model developed by Lewenstein and co-workers.

\end{abstract*}

\section{Introduction}
 The coherent manipulation of XUV light fields promises great perspectives for nonlinear spectroscopy, coherent control and high resolution metrology in the XUV spectral domain\,\cite{mukamel_multidimensional_2013, keefer_ultrafast_2023, richter_strong-field_2024, prince_coherent_2016, koll_experimental_2022, schwickert_electronic_2022, cingoz_direct_2012}. 
Most of these techniques rely on the generation of phase-stable XUV pulse sequences with precisely controlled delay. 
A prominent example is Fourier-transform spectroscopy, which has been demonstrated in the XUV domain\,\cite{cavalieri_ramsey-type_2002, wituschek_phase_2020, koll_phase-locking_2022,  jansen_spatially_2016, usenko_attosecond_2017, wituschek_tracking_2020, uhl_extreme_2022, uhl_improved_2022, de_oliveira_high-resolution_2011}. 
Related XUV frequency-comb and Ramsey spectroscopy are proposed and developed\,\cite{cingoz_direct_2012, schuster_ultraviolet_2021, dreissen_high-precision_2019} aiming at high resolution metrology. 
Other examples already demonstrated or proposed are coherent control techniques relying on two interfering XUV fields\,\cite{prince_coherent_2016,koll_experimental_2022} as well as nonlinear spectroscopy schemes mapping the coherent dynamics induced at conical intersections\,\cite{kowalewski_catching_2015, keefer_ultrafast_2023}. 

Yet, producing phase-locked XUV pulse sequences remains a technical challenge. 
Interferometry with short-wavelength radiation requires an extremely high phase stability between the interfering fields. 
This is further complicated by the fact that the highly nonlinear processes employed to generate XUV radiation can introduce detrimental phase jitter\,\cite{uhl_improved_2022}. 
In addition, it is desirable to overlap the interfering XUV fields in a collinear geometry to maximize the interference contrast. 
Yet, there is a lack of broadband optics suitable for collinear superimposition of XUV beams. 

One approach to generate collinear, phase-stable XUV pulse sequences, is the driving of high-harmonic generation (HHG) in gases with collinear infrared (IR) pulse trains. 
The advantage of this approach, is the facile control of both the timing and phase between the produced XUV pulses by controlling the IR fields. 
However, the interaction of two consecutive intense IR fields with the gas medium can cause distortions in the HHG process. 
Most prominently, the leading pulse alters the gas medium, e.g. through ionization and plasma generation, which affects the nonlinear process induced by the trailing pulse\,\cite{porat_phase-matched_2018}. 
An additional issue arises when the IR pulses overlap in time. 
The interference of the temporally overlapping IR pulses causes an amplitude modulation which impacts the HHG process in a non-trivial way. 
Phase-locked driving with two IR pulses can even lead to spectral tuning of the generated XUV radiation\,\cite{schuster_agile_2021} an effect that has been also investigated theoretically\,\cite{gulyas_oldal_generation_2020, gulyas_oldal_all-optical_2021}. 

In this work, we investigate experimentally the generation of harmonic fifteen (H15) in Ar gas by two collinearly overlapping IR pulses with the focus on producing phase-locked XUV pulse pairs for various applications. 
To this end, effects caused by temporal overlap between the driving pulses are of particular interest and are therefore in the center of our study. 
For a comprehensive mapping of the complex nonlinear response induced by two IR driving fields in HHG, we employ a special phase-modulation interferometry technique\,\cite{tekavec_wave_2006}. 
The benefits of this technique for the investigation of strong-field processes have been recently pointed out in a theoretical study\,\cite{granados_decoding_2024}. 
The experimental results are compared to two theory models, a simple perturbative approach and a non-perturbative approach based on the formalism developed by Lewenstein et al.\,\cite{lewenstein_theory_1994}. 
This comparison between experiment and theory helps to develop a mechanistic understanding of the nonlinear physics induced in rare gases by two intense driving fields, which ultimately may lead to improved strategies for realizing XUV interferometry and related coherent techniques.   

\section{Experimental method}
\begin{figure*}[htbp]
    \centering
    \includegraphics[width=0.95\textwidth]{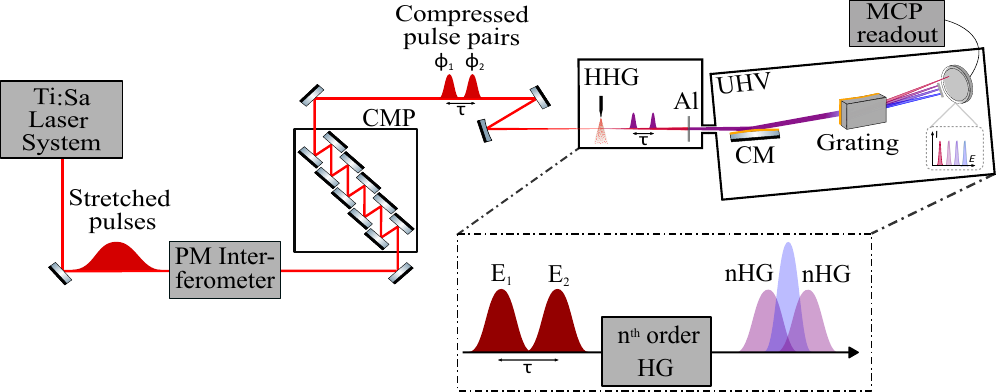}
    \caption{Experimental setup. The stretched IR pulses enter a phase-modulation interferometer producing two pulses with modulated carrier-envelope phases $\phi_1$ and $\phi_2$ and time delay $\tau$. 
    The pulses are compressed with chirped mirrors and generate high harmonics in an Ar gas jet. 
    The produced XUV radiation is analyzed with a home-built spectrometer. 
    Inset: Electric fields $E_1$ and $E_2$ generate high-order harmonics resulting in pulses at the $n\mathrm{^{th}}$ harmonic along with additional mixing contributions. 
    CMP: chirped mirror pairs, Al: aluminium filter, CM: cylindrical mirror, MCP: microchannel plate, HG: harmonic generation.}
    \label{Fig:1}
\end{figure*}

The experimental setup comprises of a standard HHG beamline combined with an optical Mach-Zehnder interferometer to drive the HHG process with two collinear IR pulses (Fig.\,\ref{Fig:1}). 
The Ti:Sapphire (Ti:Sa) laser system (Femtolasers Produktions GmbH) provides pulses centered between 790 and 800 nm with 1 kHz repetition rate and up to ~5 mJ pulse energies. 
The internal grating compressor of the laser system is detuned to stretch the pulses by a factor of $\approx 10$ in order to limit the intensity inside the optical interferometer. 
A compressor consisting of 5 pairs of chirped mirrors (HD31, Ultrafast Innovations) compresses the pulses after passing through the interferometer. 

The compressed pulses are focused ($f=250$ mm) onto an Ar gas jet produced with a nozzle (orifice diameter: 60 $\mu$m, backing pressure: 2 bar) to produce XUV pulse pairs. 
An aluminum filter (thickness: 100\,nm) is used to block the residual IR beam after the HHG process. A cylindrical mirror (CM) enhances the divergence of the XUV beam for a better illumination of the grating inside our home-built XUV spectrometer. The harmonic spectrum is recorded by a microchannel plate (MCP), a phosphor screen and a camera. 
To simplify the optical response from the nonlinear conversion in the Ar gas jet, a spatial mask is inserted behind the XUV grating in order to transmit only H15 onto the MCP detector. 
The XUV radiation generates a photocurrent in the MCPs which is the relevant signal detected in the experiment. 

\begin{figure}[htbp]
    \centering
    \includegraphics[width=0.5\columnwidth]{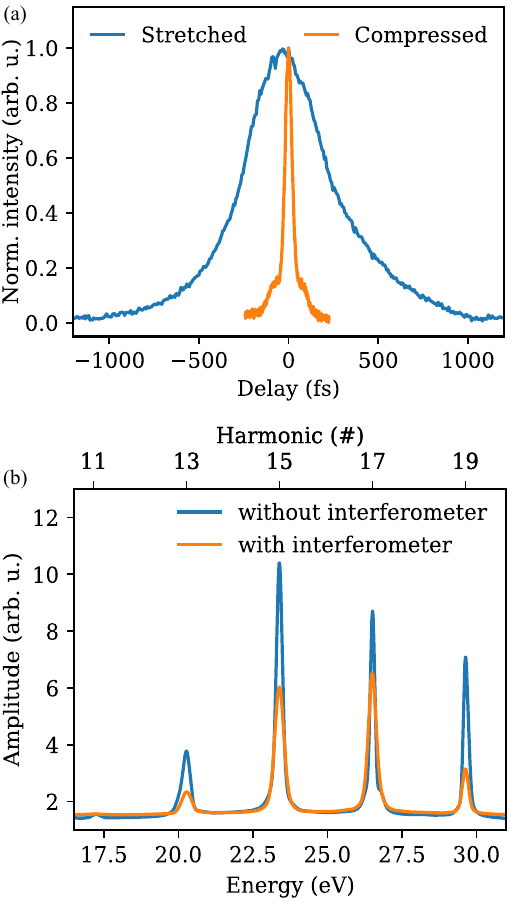}
    \caption{(a) Intensity autocorrelation traces of stretched input pulses (blue) and pulses compressed with the chirped mirror pairs (orange), (b) harmonic spectra recorded without (blue) and with (orange) the phase-modulated interferometer inserted in the beamline.}
    \label{fig:2}
\end{figure}

The temporal profile of the IR pulses was characterized by second-harmonic intensity autocorrelation (AC) before entering the interferometer and after passing through the chirped-mirror compressor (Fig.\,\ref{fig:2}a), yielding pulse widths of 327\,fs and 30\,fs, respectively. 
Type I second harmonic generation in a BBO crystal of 10\,\textmu m thickness was used for the characterization.  
The visible shoulders in the AC traces of the re-compressed pulses imply residual third-order dispersion, which is not observed in the laser output when the pulses are fully compressed with the internal grating compressor of the laser. 

Figure \ref{fig:2}b shows the harmonic spectrum (integrated for 0.25\,s) produced without (blue) and with the interferometer (orange) placed in the beam path. 
Here, one arm of the interferometer was blocked to avoid interference. 
Both HHG spectra are recorded for a pulse energy of 300\,\textmu J entering the vacuum setup. 
In both cases, the harmonic spectra span up to around 30\,eV, which clearly consists of peaks corresponding to harmonics 13 - 19. 
However, the harmonic yield is reduced by a factor of at least 1.7 when the interferometer is inserted into the beamline presumably due to the not fully compensated dispersion introduced by the interferometer. 

The details of the interferometer have been described elsewhere\,\cite{ganeshamandiram_phase-modulation_2025-1}. 
We employ a special phase-modulation (PM) technique\,\cite{tekavec_wave_2006} which has proven to substantially increase the sensitivity of interferometric measurements\,\cite{tekavec_wave_2006, bruder_phase-modulated_2015, bruder_efficient_2015, jana_fluorescence-detected_2024}. 
This is important to provide sufficient dynamic range for a comprehensive characterization of the highly nonlinear processes induced in the HHG process. 
For details about the phase modulation technique, we refer to Ref.\,\cite{tekavec_wave_2006}. 
Briefly, an acousto-optical modulator (AOM) is placed in each arm of the Mach-Zehnder interferometer. 
Phase-locked driving of the AOMs at distinct radio-frequencies $\Omega_{1,2}\approx 110\,$MHz leads to a modulation of the interference signal at a clean difference frequency $\mathrm{\Omega_{21}=\Omega_2 - \Omega_1}=15\,$Hz. 
This modulation and harmonics thereof are imprinted into the detected photocurrent generated in the MCPs by the XUV radiation. 
The photocurrent is amplified with a home-built fast amplifier and fed into the current input of a commercial lock-in amplifier (MFLI, Zurich Instruments) for demodulation. 
Care is taken to avoid saturation in the amplification which would cause spurious nonlinear signal contributions. 
At the same time, a portion of the interferometer output is spectrally filtered in a home-built monochromator and used as reference signal for the lock-in amplification. 
This phase-synchronous detection using an optical reference reduces the phase noise in the reconstructed interferograms and provides additional phase information\,\cite{tekavec_wave_2006}. 

Since the H15 is optically selected in the experiment, one would intuitively expect a clean modulation of the photocurrent at $15 \times \Omega_{21} = 225\,$Hz. 
However, the temporal overlap of the driving pulses leads to a nonlinear mixing of the IR fields in the HHG process, resulting in a complex distribution of many nonlinear signal contributions even if a single harmonic of the produced XUV radiation is optically selected. 
Moreover, the amplitudes of these signal contributions span over many orders of magnitude, making the full characterization of the nonlinear HHG process challenging. 
Yet, with the lock-in detection, the photocurrent signal is conveniently disentangled into separate detection channels based on the individual phase-modulation signature of each signal contribution, as further outlined below. 
This enables optimal amplification of each signal contribution and advanced signal recovery capabilities. 

\section{Theoretical description of the nonlinear light-matter interaction in the gas medium}
Below, we outline the theoretical modeling of the HHG process driven by two temporally overlapping pulses. 

\subsection{Driving field}
The Ar atoms in the gas jet interact with the electric fields 
\begin{equation} \label{eq:E}
    E_j=\epsilon_j(t)\exp{(i\omega_0 t +\phi_j) },
\end{equation} 
of the two IR pulses (j=1,2), where $\epsilon_j$ denotes a Gaussian envelope function, $\omega_0$ the carrier frequency of the IR fields and $\phi_j=\Omega_j t$ their carrier-envelope phase which is modulated by the AOMs. Higher order phase terms are neglected for simplicity. 
The total electric field is thus
\begin{equation}\label{eq.E}
    E(t) = E_1(t) + E_2(t-\tau), 
\end{equation}
where $\tau$ denotes the temporal delay between both pulses. 
The intensity of the combined fields is
\begin{equation}\label{eq:Idrv}
    I(\tau, t)=\epsilon_{CC}(\tau)(1+\cos[\omega_0 \tau + \Omega_{21}t]),
\end{equation}
where $\epsilon_{CC}$ denotes a Gaussian envelope function, whose width corresponds to the cross-correlation width of the two IR pulses. 

Driving the Ar gas medium with the intense IR pulses leads to emission of XUV radiation at odd harmonics of the frequency of the driving field. 
The interference between the two driving fields and their nonlinear mixing in the HHG process leads to a complex response for temporally overlapping driving pulses. 
As a result, the emitted harmonics from the gas medium each contain contributions with phase terms of the form
\begin{eqnarray} \label{Eq: phase}
    \theta_{n,q} = q (\omega_0 \tau + \Omega_{21}t)\, ,  
\end{eqnarray}
where $n \in N$ denotes the order of the emitted harmonic and $q \in N$ labels what we call in the remainder the $q$-order phase contribution to the $n\mathrm{^{th}}$ harmonic emission. 

This behavior was previously described in the framework of perturbation theory for second-harmonic generation ($n=2$)\,\cite{bruder_phase-modulated_2017}, hence for a low-order nonlinear process. 
For the special case of high-gain harmonic generation (HGHG) in seeded free-electron lasers, the nonlinear mixing of temporally overlapping driving pulses was investigated for $n=6$\,\cite{wituschek_high-gain_2020}. 
In contrast, the current work investigates high-order nonlinear processes in atomic gases for $n=15$ and compares the perturbative and non-perturbative description of the nonlinear light-matter interaction driven by two collinear IR fields. 

\subsection{Perturbative model}
We first describe the light-matter response by perturbation theory. Although this approach over-simplifies the nonlinear HHG process, the model is instructive and helps to develop an intuitive understanding for the appearance of the $q$-order phase terms. 
In the experiment, the intensity $I_{15}$ of H15 produced in the Ar gas is detected. In the perturbative approach, it is
\begin{eqnarray} \label{Eq: Perturb1}
    I_{15} &\propto&\chi^{(15)}|E^{15}|^2 \notag \\
                    &=& \chi^{(15)}|\sum_{k=0}^{15} \binom{15}{k}\, E_1^{\, 15-k}(t)\,E_2^{\,k}(t-\tau)|^2 ,
\end{eqnarray}
where we used Eq.\,\ref{eq.E} and $\chi^{(15)}$ is the $\mathrm{15^{th}}$ order of the perturbative expansion of the susceptibility of the Ar gas. 
This can be simplified to 
\begin{equation}\label{Eq: Perturb2}
    I_{15}(\tau) \propto \mathrm{const.} + \sum_{q=0}^{15} C_q(\tau) \cos(\theta_{n,q}).
\end{equation}
$I_{15}$ corresponds to the $\mathrm{15^{th}}$ order interferometric autocorrelation of the two IR pulses and can be decomposed into the $q$-order phase contributions oscillating at harmonic frequencies $q\omega_0$ w.r.t. $\tau$ and $q \Omega_{21}$ w.r.t. the laboratory time $t$. 
$C_q$ describes the Gaussian envelope of the $q$-order contribution. 

\begin{figure}[htbp]
    \centering
    \includegraphics[width=0.5\columnwidth]{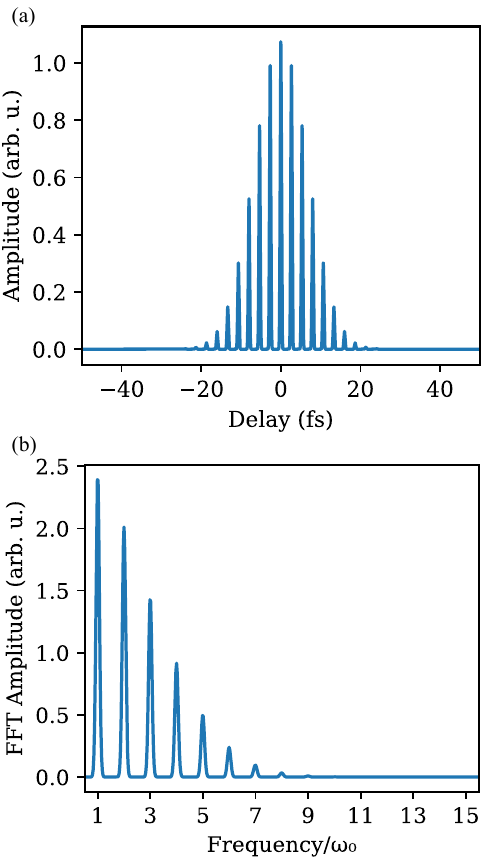}
    \caption{H15 generation calculated with perturbation theory for two IR driving pulses. (a) time domain interferogram of the produced H15 radiation and (b) it's Fourier transform.}
    \label{fig:3}
\end{figure}
We assume IR pulses with Gaussian electric field envelope of 30\,fs FWHM. 
Fig.\,\ref{fig:3} shows the calculated signal in the time domain (a) which corresponds essentially to the $\mathrm{15^{th}}$ order interferometric autocorrelation of the pulses. 
Its Fourier transform w.r.t. $\tau$ (b) clearly shows the composition into discrete  $q$-order contributions, whose amplitudes rapidly decrease with increasing $q$, demanding high dynamic range for a complete mapping of the XUV field produced by the two overlapping IR pulses.\\

\subsection{Non-perturbative model}
While the above perturbative model is well suited to describe low-harmonic generation in bulk crystals and gases, the physics of HHG in a gas target requires generally more involved models.  

The process for HHG is commonly described by the well-established three step model\,\cite{corkum_plasma_1993,lewenstein_theory_1994}. When an intense IR laser field interacts with a suitable gas target, an electron is emitted through tunnel ionization. The emitted electron is accelerated by the oscillating electric field and re-collides with the parent ion when the polarity of the field reverses. The re-collision leads to the emission of harmonics with odd integer multiple photon energies. 

For a thorough calculation of this process we adopt the theoretical approach developed by Lewenstein \textit{et. al} \cite{lewenstein_theory_1994} using the field described in Eq.\,\ref{eq.E}, but omitting the $\Omega_{21}$-modulation for simplicity. 

To obtain the emitted XUV field, the nonlinear polarization and thus the induced nonlinear dipole moment $x(t)$ is obtained from the solution of the time-dependent Schrödinger equation (TDSE) using the strong-field approximation (SFA) \cite{lewenstein_theory_1994, sansone_nonadiabatic_2004,sansone-pra-2009}: 

\begin{equation}\label{Eq: Lewenstein1}
\begin{aligned}
    x(t;\tau) = i\int_{-\infty}^{t}\,dt'\left[\frac{\pi}{\zeta+i(t-t')/2}\right]^{3/2}\,E(t';\tau)\,d^*_x[p_s-A(t;\tau)]\,d_x[p_s-A(t;\tau)] \\
    \times\, e^{-iS(p_s,t,t';\tau)}\,dt'\, +\,c.c. ,
\end{aligned}
\end{equation}

where $\zeta$ is a positive regularization constant, $A$ indicates the vector potential associated to the IR field, $d_x$ is the dipole matrix element between the ground state and a state in the continuum, and $p_s$ indicates the stationary momentum:
\begin{equation}
    p_s=\frac{1}{t-t'}\int_{t'}^tA(t'';\tau)dt''
\end{equation}

The phase
\begin{equation}\label{Eq: Action}
   S(p_s,t,t';\tau) = \int_{t'}^{t}\,dt''\,\left(\dfrac{[p_s-A(t'';\tau)]^2}{2}+I_p\right)
\end{equation}
is the quasiclassical action. In Eq.\,\ref{Eq: Action}, $I_p$ stands for the ionization potential of the gas target. 
Note, that in the special case where we drive the harmonics with two laser fields, the electric field is defined as the sum of two individual fields Eq.\,\ref{eq.E}. 

Similarly, the vector potential $A(t)$, the quasiclassical action $S$, and finally the time-domain dipole moment $x(t)$ are calculated considering the sum of two electric fields. 
Fig.\,\ref{fig:4}a shows a flow chart of the most important calculation steps: 
\begin{figure*}[htbp]
    \centering
    \includegraphics[width=1\columnwidth]{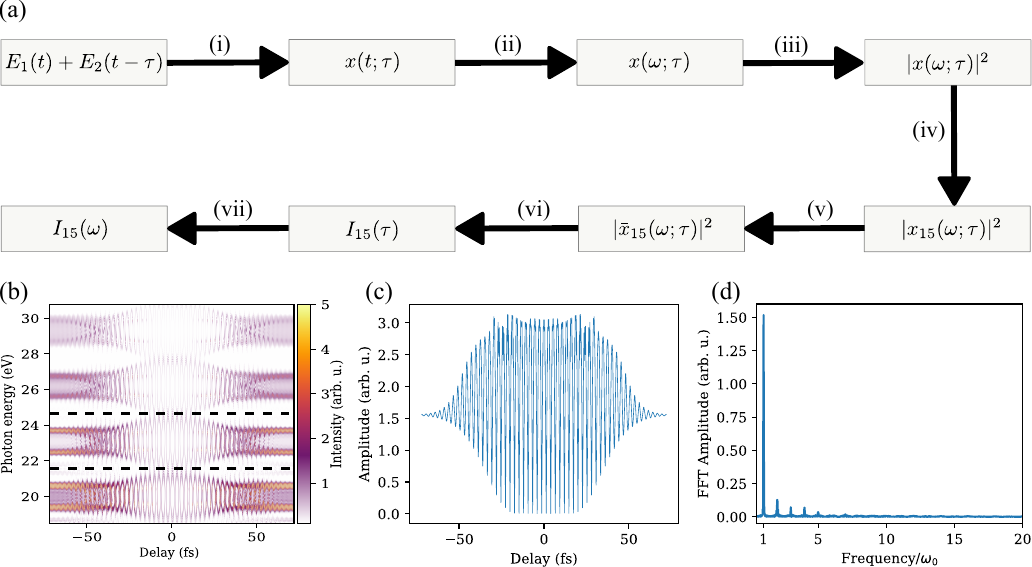}
    \caption{Calculation steps in the non-perturbative model. (a) Flow chart. (b) Calculated $x(\omega ; \tau)$, dashed lines indicate the boundaries of the band pass filter to select H15. (c) $I_{15}(\tau)$, (d) $I_{15}(\omega)$ signals.}
    \label{fig:4}
\end{figure*}

(i) the calculation of the nonlinear dipole moment $x(t;\tau)$ in the time domain as a function of inter-pulse delay $\mathrm{\tau}$ starting from the sum of two electric fields, each with a peak intensity of $\mathrm{2.456\times 10^{14}\,W/cm^2}$. Note that the upper limit of the integration in Eq.\,\ref{Eq: Lewenstein1} is set to $0.65$ times the optical period of the IR field, which restricts the simulations solely to the short electron trajectories. 

(ii) Fourier transform w.r.t. the laboratory time $t$ yielding the dipole moment in the frequency domain $x(\omega;\tau)$. 

(iii) The modulus square of $x(\omega;\tau)$ is proportional to the emitted XUV spectrum as a function of the pulse delay $\tau$. The data is plotted in Fig.\,\ref{fig:4}b. 

(iv) A band-pass frequency filter (dashed horizontal lines in Fig. \ref{fig:4}b) selects the H15 spectral component, denoted $|x_{15}(\omega; \tau)|^2$.

(v) We compute $\left|x_{15}(\omega; \tau)\right|^2$ for IR intensities ranging from $\mathrm{2.947\times 10^{13}\,W/cm^2}$ in 9 specific steps to $\mathrm{2.456\times 10^{14}\,W/cm^2}$ and average $|x_{15}(\omega; \tau)|^2$, yielding $|\bar{x}_{15}(\omega; \tau)|^2$. 
The intensity steps and their weighting factor were chosen according to a 3D focal volume average of a Gaussian TEM$_{00}$ mode as described in Ref.\cite{wiese_strong-field_2019}.  

(vi) Integrating this data along $\omega$ yields the signal proportional to the XUV intensity of H15 $I_{15}(\tau)\propto \int_{-\infty}^\infty|\bar{x}_{15}(\omega, \tau)|^2$, as detected in the experiment.  This data is shown in Fig. \ref{fig:4}c. 

(vii) Fourier transform of this signal w.r.t. $\tau$  decomposes the $I_{15}$ into the $q$-order components (Fig.\,\ref{fig:4}d), in analogy to the analysis performed for the perturbative model and the experimental data.
\begin{figure}[htbp]
    \centering
    \includegraphics[width=0.5\columnwidth]{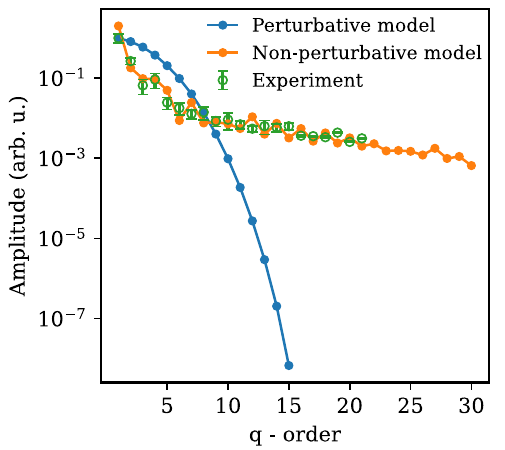}
    \caption{Amplitude comparison: the peak amplitudes of the $q$-order signal components are shown for the experimental data (green), the perturbative (blue) and the non-perturbative model (orange). For $q \leq 15$, experimental error bars reflect the fluctuations between three consecutive experiments and for $q>15$ the error of the fitting routine for a single data set.}
    \label{fig:5}
    \end{figure}
\begin{figure*}[htbp]
    \centering
    \includegraphics[width=0.8\textwidth]{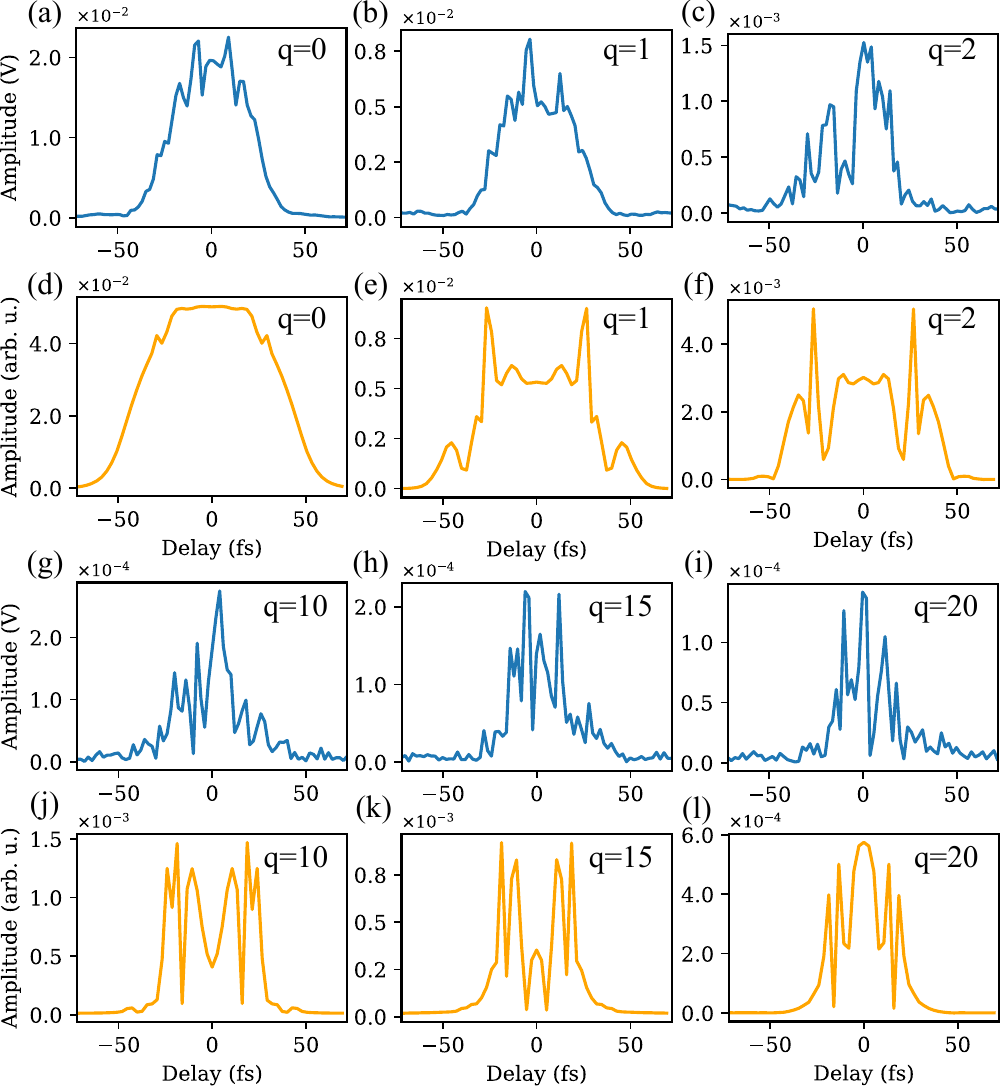}
    \caption{Comparison of experimental (blue) and calculated (orange) $I_{15}(\tau)$ time-domain signals as a function of the delay $\tau$ decomposed into individual $q$-order contributions. Only the envelope of the signals are shown (see text).}
    \label{fig:6}
\end{figure*}
\begin{figure*}[htbp]
    \centering
    \includegraphics[width=0.8\textwidth]{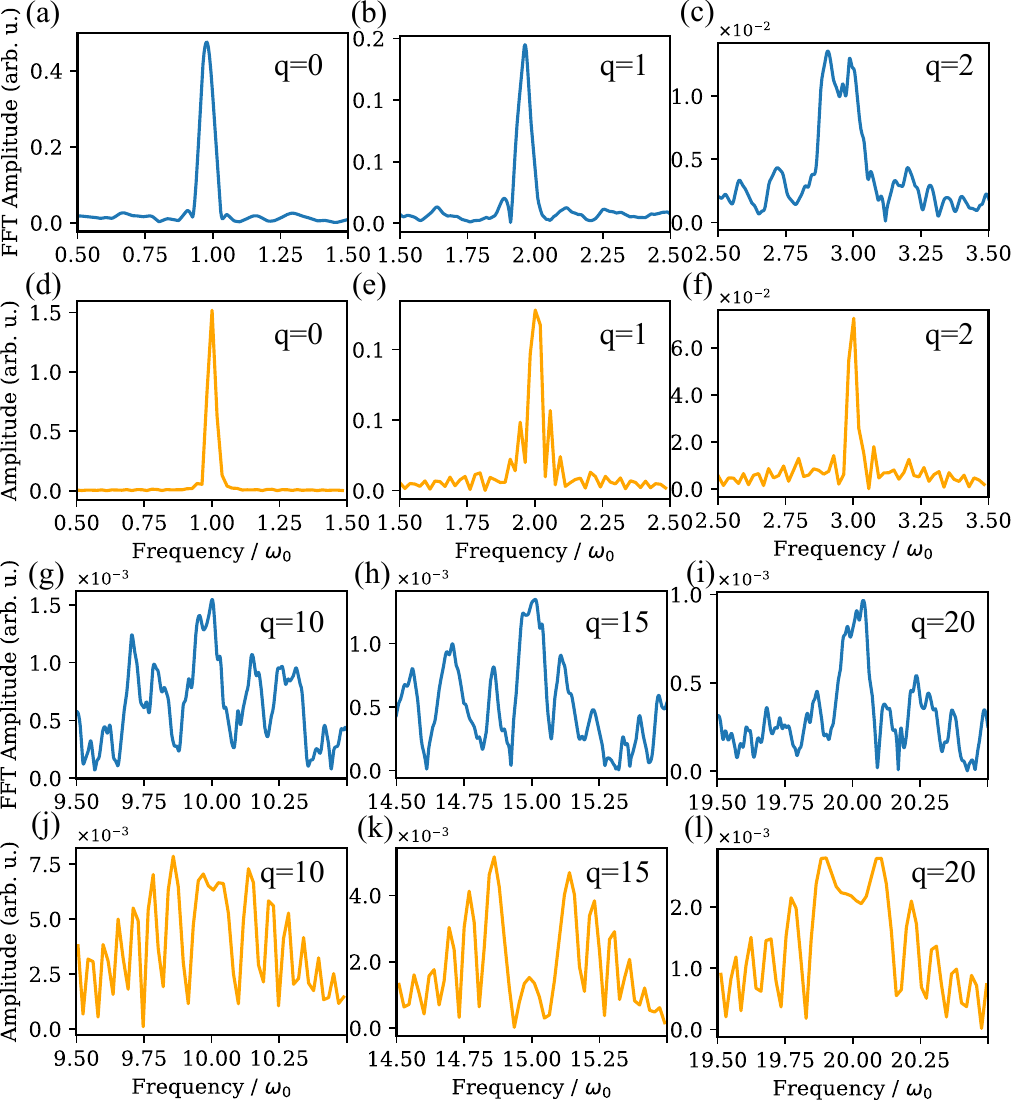}
    \caption{Corresponding Fourier transforms of $I_{15}(\tau)$ for the same signal components as shown in Fig.\,\ref{fig:6}.}
    \label{fig:7}
\end{figure*}

\section{Results}
The experimental data was recorded for H15 of the emitted XUV radiation and a $\tau$-delay range of $-200$ to $+200$\,fs incremented in 2\,fs steps. 
For each $\tau$ step, we detected the signal for 1\,ms and demodulated it simultaneously with three lock-in channels. 
Measurements were repeated for the channels tuned to different $q \Omega_{21}$ frequencies to map the full light-matter response for a range of $q=1 - 21$. 
For the theoretical modeling the $\tau$-delay range was $-72$ to $+72$\,fs and $\tau$-increments were ~9\,as and Gaussian electric fields with a FWHM of 30\,fs were assumed. 
Note, that the theoretical models omit for simplicity the carrier-envelope phase modulation induced in the experiment and the decomposition of the $q$-order phase terms relies solely on the $q \omega_0$-dependence of the signals.
This approximation seems justified given the rather long IR pulses used in the experiment, for which carrier-envelope phase effects should be negligible in the HHG process. 

For each $q$-order contribution of $I_{15}$, we evaluated the peak amplitude by fitting the experimental and theoretical data with a Gaussian function. Note that a bootstrapping technique with 95$\%$ confidence interval was applied on the experimental data prior to the Gaussian fits, to reduce the noise and make the amplitude extraction more reliable.
The comparison between the experimental data, the perturbative and the non-perturbative model is shown in Fig.\,\ref{fig:5}. To this end, we scaled the experimental data and the perturbative model in order to have an amplitude of one for $q=1$. The global scaling factor for the non-perturbative model was determined upon a least-square fit to the experimental data. 

In the perturbative model, the amplitudes decrease strongly with increasing $q$, by $>8$ (8.17) orders of magnitude between $q=1$ and $q=15$. 
There is no signal contribution for $q>15$, which is intuitively expected given that we essentially detect the $\mathrm{15^{th}}$ harmonic of the driving field. 

In contrast, the experimental data shows two distinct differences: 
the amplitude decrease follows clearly a different functional form than predicted by the perturbative model, and signal contributions for up to $q=21$ were detected well discernible from the noise in the measurements. 
In contrast, both of these observations are in good agreement with the non-perturbative calculations, where we find a good prediction of the experimentally observed amplitudes including the extension of signal contributions beyond $q=15$. 

This result clearly confirms that the high-order nonlinear light matter interaction in the gas target cannot be well described by perturbation theory, in contrast to low-order harmonic generation. 
Moreover, the findings confirm, that the non-perturbative calculations based on the theoretical framework by Lewenstein et. al\,\cite{lewenstein_theory_1994} provide an adequate description of the complex nonlinear response when induced by two interfering, temporally overlapping IR fields.

For a more detailed comparison of the experimental data and the non-perturbative model, we show in Fig.\,\ref{fig:6} and Fig.\,\ref{fig:7}, a comparison of the time and frequency domain data for a few selected $q$-order signal contributions. 
The detected $q$-order contributions of $I_{15}$ are naturally modulated at high frequency w.r.t. $\tau$. 
For a more convenient representation and analysis of the signals, we plot the envelope of each $q$-order contribution. 
For the experimental data this is readily achieved by decomposing the complex-valued signal output of the lock-in amplifier $R \exp[i \phi]$ into amplitude $R$ and phase $\phi$ and plot $R$. 
the same decomposition into amplitude and phase is applied to the calculated data. 
Note, that in the experimental data the frequency axis is slightly miscalibrated due to experimental uncertainties, causing the slight systematic red-shift in the experimental spectra.

In Fig.\,\ref{fig:6} and \ref{fig:7}, the signal amplitudes exhibit distinct modulations, well discernible from the noise. 
For some $q$-orders the structure of these modulations matches qualitatively between the experiment and theory. 
For other $q$-orders clear deviations are visible. 
In contrast, the perturbative model does not predict any of these modulations (not shown). 
This confirms the indeed complex behavior of the nonlinear light-matter response and the fact, that the current non-perturbative model cannot capture all details of the experiment. 

Moreover, we note that between $q=1$ and $q=21$ the signal amplitude decreases by roughly two orders of magnitude, requiring high dynamic range to detect all contributions with sufficient vertical resolution. 
While achieving a dynamic range of $> 100:1$ is usually not a problem with modern electronic circuits and detectors, we find that the signal amplitude of the $q > 11$ channels is already at a level comparable to the noise floor of the $q=1$ channel. 

Hence, the higher $q>11$-order signal contributions to $I_{15}$ would not be detectable with a standard signal processing scheme, rendering an incomplete characterization of the nonlinear light-matter response. 
In contrast, the detection of the $q$-order signal contributions in separate lock-in detection channels improves the signal-to-noise for the detection of the higher-order signal components. 
Similar behavior has been reported for the detection of nonlinear signals stemming from many-body interactions in dilute systems\,\cite{bruder_delocalized_2019}. 

\section{Discussion and conclusions}
The presented study provides useful implications for the development of coherent nonlinear spectroscopies and coherent control experiments using XUV and potentially soft X-ray radiation. 
We focused here on the technical approach using multiple collinear IR pulses driving HHG in an atomic gas target. 
The nonlinear mixing of the two driving fields in the HHG process results in a highly complex light-matter response whose mapping requires detection with high dynamic range. 

Our results imply that the theoretical framework developed by Lewenstein et al.\,\cite{lewenstein_theory_1994} provides an adequate description for the strong-field light matter response, albeit quantitative agreement between experiment and theory could not be achieved for all analyzed features. 
This provides a basis for further experimental and theoretical investigations. 

In potential applications, such as XUV Fourier transform spectroscopy and related techniques, the spectroscopic information is deduced from a Fourier analysis. 
Our results imply, that these Fourier spectra can be contaminated by $q\omega_0$-frequency contributions if the driving fields temporally overlap. 
While these signals can be clearly identified and filtered in the Fourier domain, their large amplitudes may dominate or even completely mask the genuine spectral features of interest. 
As the parasitic $q\omega_0$-signal contributions appear only during temporal overlap between the driving pulses, short IR pulses will improve the situation, limiting the spurious signals to a short delay range which may be omitted in the data acquisition without losing relevant information. 

\begin{backmatter}
\bmsection{Funding}
The authors acknowledge the funding from Deutsche Forschungsgemeinschaft RTG 2717, STI 125/24-1 and DFG SA3470/3-1, and European Research Council Starting Grant MULTIPLEX (101078689). This research is part of the COST Action NEXT (CA22148).

\bmsection{Acknowledgment}

\bmsection{Disclosures}
The authors declare no conflicts of interest.

\bmsection{Data Availability Statement}
The experimental and simulation data included in this work are available on the open repository: \textit{Accession codes will be available before publication}.
\end{backmatter}

\bibliography{PM-XUV_Paper, Bruder, 2026intHHG}

\end{document}